\documentclass[aps,prl,reprint,superscriptaddress,floatfix,nofootinbib]{revtex4-2}
\usepackage[T1]{fontenc}
\usepackage[utf8]{inputenc}
\usepackage{graphicx}
\usepackage{booktabs}
\usepackage{array}
\usepackage{amsmath,amssymb,amsfonts,mathtools}
\usepackage{bm}
\usepackage{longtable}
\usepackage{multirow}
\usepackage{dcolumn}
\usepackage{xcolor}
\usepackage{esint}
\usepackage{comment}
\usepackage[hidelinks]{hyperref}
\usepackage{titlesec}
\renewcommand\thesection{\Roman{section}}
\titleformat{\section}{\normalfont\large\bfseries}{\thesection.}{0.6em}{}

\newcommand{\Aext}{\mathbf{A}_{\mathrm{ext}}}

\newcommand{\TDGL}{TDGL}

\newcommand{\safeincludegraphics}[2][]{%
  \IfFileExists{#2}{%
    \includegraphics[#1]{#2}%
  }{%
    \fbox{\parbox[c][0.12\textheight][c]{0.9\linewidth}{\centering Missing figure file:\\[0.5ex]\texttt{#2}}}%
  }%
}

\begin{document}

\title{On the Detection of Curl-Free Gauge Fields}

\author{Armen Gulian}
\email{gulian@chapman.edu}
\affiliation{Laboratory of Advanced Quantum Materials and Devices, Institute for Quantum Studies, Chapman University, Orange, CA 92866, USA}

\author{Will Benston}
\affiliation{Laboratory of Advanced Quantum Materials and Devices, Institute for Quantum Studies, Chapman University, Orange, CA 92866, USA}

\author{Vahan Nikoghosyan}
\affiliation{Laboratory of Advanced Quantum Materials and Devices, Institute for Quantum Studies, Chapman University, Orange, CA 92866, USA}
\affiliation{Institute for Physical Research, National Academy of Sciences, Ashtarak 0304, Republic of Armenia}

\begin{abstract}
In quantum theory, electromagnetic gauge fields enter directly into the phase evolution of the wavefunction and can even influence quantum systems in regions where the associated electric and magnetic fields vanish. The Aharonov–Bohm effect demonstrates that such gauge fields produce observable consequences when a coherent quantum system encloses magnetic flux along a doubly connected path. This has led to the widespread view that curl-free gauge fields are undetectable in simply connected systems, giving rise to a form of topological blindness. Here we show that this conclusion is not fundamental. During nonequilibrium processes, collective quantum systems can develop transient responses to curl-free gauge fields without enclosing magnetic flux in a static geometry. Using superconducting condensates as a concrete example, we demonstrate that the evolving phase of the macroscopic wavefunction generates supercurrents and voltage pulses whose time integral is proportional to the open-path line integral of the vector potential. In contrast to the conventional Aharonov–Bohm effect, the resulting response is not restricted modulo the flux quantum and may greatly exceed the scale associated with static doubly connected geometries. The mechanism can be interpreted as a dynamical closure of the gauge contour in spacetime and is supported by gauge-invariant arguments, time-dependent Ginzburg–Landau theory modeling, and numerical simulations. These results establish a general principle for detecting curl-free gauge fields and suggest new approaches for probing hidden gauge structures in quantum matter and beyond.
\end{abstract}


\maketitle

The role of electromagnetic potentials in quantum physics represents one of the most profound departures from classical intuition. In classical electrodynamics, physical effects are fully determined by electric $\mathbf{E}$ and magnetic $\mathbf{H}$ fields, while scalar $\varphi$ and vector $\mathbf{A}$ potentials serve as auxiliary mathematical constructs. In quantum mechanics, however, potentials $\varphi$ and $\mathbf{A}$ acquire direct physical significance. This conceptual shift is most profoundly illustrated by the Aharonov--Bohm (AB) effect~\cite{Aharonov1959}, where charged particles respond to electromagnetic potentials even in regions where vanish both electric ($\mathbf{E}=\nabla\varphi$ - $\partial\mathbf{A}/\partial {t} \equiv0$) and magnetic ($\mathbf{H}=\nabla\times\mathbf{A} \equiv0$) fields \cite{Ehrenberg1949}. (From now-on, we use theoretical units $2e=\hbar=c=k_B=1$.) 

In the classical world, a curl-free vector potential can always be eliminated by a gauge transformation (accompanied by a corresponding transformation of the scalar potential). In the presence of charged quantum objects, however, the situation is drastically different.
The phase $\theta$ of the wavefunction
\begin{equation}
\psi=|\psi|e^{i\theta}
\label{eq:psi}
\end{equation}
transforms together with the gauge field and thereby retains the physical influence of the curl-free vector potential. Nevertheless, it was considered in the community, that this accumulation has no physical effects unless the quantum object is doubly or multiply connected and encircles the magnetic flux which generates this curl-free field \cite{Tonomura1998}. In case of point quantum object, such as an electron, the quantum trajectories should be closed (see e.g. \cite{Feynman1964,Aharonov2005}). If the trajectory is not closed, or if the flux is not completely enclosed, the effect of the curl-free vector potential has traditionally been regarded as unobservable (topologically prohibited) \cite{Yang1996,Belot1998,Healey2007,Nounou2010,Gomes2025}. 

In this article we show that this conclusion is wrong in general.

We begin with a gedanken experiment which will explain why the topological blindness of singly connected quantum sensors is not a fundamental property. Imagine an experimental situation plotted in Fig. 1. The ideal source of the curl-free vector potential is the infinite solenoid, in which case, for a steady current $I$, both electric and magnetic fields vanish. However, the vector potential remains finite and exhibits a characteristic spatial dependence outside the solenoid:
\begin{figure}[t]
\centering
\safeincludegraphics[width=0.75\columnwidth]{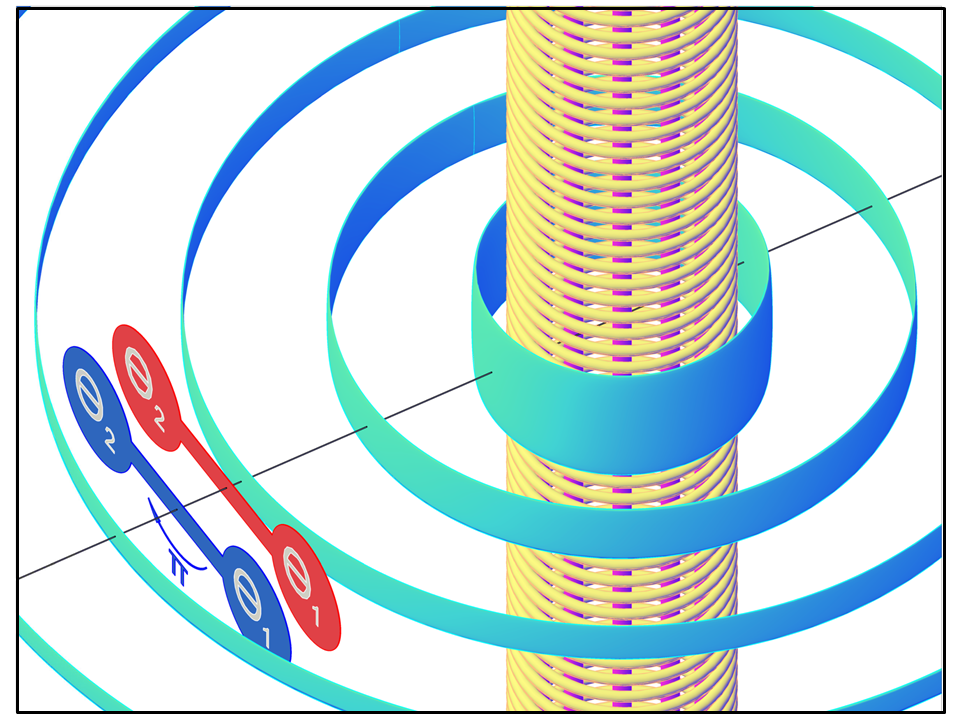}
\caption{Schematic of the \textit{gedanken} experiment. Concentric rings correspond to the azimuthal component $A_\theta$. Solenoid windings are depicted in a way to expose the magnetic flux lines. Flat "barbells" are explained in the main text.}
\label{fig:solenoid}
\end{figure}
\begin{equation}
A_\theta(r)=\ \frac{a}{r},   ~~a={{\mu_0}{nIR^2}}/{2}
\label{eq:AthetaSI}
\end{equation}
where $A_\theta$ is the azimuthal component of the vector potential, $r$ is the distance from the coil axis, $\mu_0$ is the vacuum permeability,  $R$ is the solenoid radius, and $n$ is the winding density. 

Two flat "barbells" are the superconducting strips most important for our experiment . If a superconducting strip is placed along an open contour parallel to the azimuthal direction of the curl-free field, no current can flow in the stationary case. The corresponding supercurrent density is
\begin{equation}
\mathbf{j}_s=|\Psi|^2(\nabla\theta-\mathbf{A}).
\label{eq:js_basic}
\end{equation}
Integrating Eq.~(\ref{eq:js_basic}) along such an open superconducting segment and imposing $\mathbf{j}_s=0$ shows that the condensate phase adjusts so as to compensate the vector potential. In this sense, the absence of current is not due to irrelevance of the field $\mathbf{A}$, but to the adjustment of the superconducting phase.
Each strip acquires a phase difference between its ends that compensates for the action of the line integral of $\mathbf{A}$, thus constituting a phase battery. (For a closed superconducting loop surrounding the solenoid the situation is different. The single-valuedness of the condensate wavefunction imposes a quantization condition, and one obtains the usual fluxoid relation.) 

If the strips are sufficiently close, Josephson tunneling can occur between the nearby ends~\cite{Josephson1962}. In the original orientation, the two endpoint phase batteries compensate each other and no tunneling current flows. However, if one strip is rapidly rotated by $\pi$-angle (as shown in Fig. 1) and the Josephson contact is established before phase memory is lost, the junction compares phases belonging to different moments in the condensate history. A transient current spike results, and then decays as the phase relaxes.

 The physical lesson here is clear: \textit{sensitivity to a curl-free quantum field does not require a static doubly connected spatial geometry, provided the protocol itself stores and compares dynamic phase information}.
 We can go now beyond this gedanken experiment  considering the adjustment of the superconducting phase itself as a dynamic process.

Our second and more direct nonequilibrium route to observing the curl-free field  $\textbf{A}_{ext}$ is related to the Kibble--Zurek mechanism~\cite{Kibble1976,Zurek1985,Zurek1996}. This mechanism has motivated in the past successful experimental searches for spontaneous magnetic-flux nucleation during rapid cooling of superconducting films through $T_c$~\cite{Carmi1999}. In Type II superconductors, the relevant topological defects are vortex states or vortex lines, whose phenomenology is reviewed in~\cite{Maniv2001}. Near the transition, critical slowing down freezes phase correlations over a finite length. The initial phase field is therefore generically random on short scales, with zero ensemble-averaged gradient, while the equilibrium state in the presence of a static external field requires
\begin{equation}
\nabla\theta=\Aext.
\label{eq:equilibrium_phase}
\end{equation}
Immediately after nucleation, the supercurrent therefore has a nonzero expectation value,
\begin{equation}
\langle \mathbf{j}_s(0)\rangle\propto -\Aext,
\end{equation}
which relaxes to zero as the condensate reorganizes its phase. The characteristic timescale is the Ginzburg--Landau relaxation time,
\begin{equation}
\tau_{GL}=\frac{\pi}{8(T_c-T)},
\label{eq:tauGL}
\end{equation}
in our adopted units.

The gauge-invariant endpoint phase difference across the strip (consider a single "barbell" in Fig.~1) is
\begin{equation}
\delta(t)=\Delta\theta(t)-\int_{\mathrm{strip}}\mathbf{A}\cdot d\mathbf{l},
\label{eq:delta_main}
\end{equation}
and the measurable voltage  is simply
\begin{equation}
V(t)=\dot\delta(t).
\label{eq:Vdelta}
\end{equation}
If the phase is initially unadjusted, $\Delta\theta(0)\approx 0$, while at long times the phase approaches the equilibrium value corresponding to Eq.~(\ref{eq:equilibrium_phase}). Integrating Eq.~(\ref{eq:Vdelta}) then yields a central sum rule:
\begin{equation}
\int_0^{\infty}V(t)dt=\int_{\mathrm{strip}}\mathbf{A}_{ext}\cdot d\mathbf{l}=\frac{\Theta_{az}}{2\pi}\,\phi,
\label{eq:sumrule}
\end{equation}
where $\Theta_{az}=L/r$ is the azimuthal angle subtended by a strip of arc length $L$ at radius $r$.

Equation~\eqref{eq:sumrule} is our final result. It is gauge invariant because it is derived from the endpoint phase difference $\delta$ (and not from a bare open-path integral taken in isolation). It makes the physical content transparent: in a nucleation experiment the system responds to the absolute line integral of the curl-free field, rather than only to flux modulo the flux quantum. This is precisely what makes the open, dynamical geometry attractive for sensing applications and for probing the underlying gauge structure (see \textit{Appendix A} in End Matter). 

The gedanken and the nucleation voltage spike experiments can be understood in modern gauge-theoretic language through holonomy. The relevant gauge-invariant quantity is the Wilson loop \cite{Wilson1974}
(see also \cite{Wen2004,Xiao2010}):
\begin{equation}
W(\mathcal{C})=\exp\left(i\oint_{\mathcal{C}}A_\mu dx^\mu\right),
\label{eq:wilson}
\end{equation}
where $A_\mu=(\varphi,-\mathbf{A})$ is the electromagnetic four-potential. In this context, observability means non-zero values of the Wilson loop. The experimental protocols of the described above construct precisely the closed contours needed for a gauge-invariant phase comparison. The closure is supplied dynamically in spacetime rather than statically in space.

This resolves the apparent contradiction with topological blindness. Here, Wilson loops are used  as an explanatory framework, not as a starting postulate of the physical mechanism. A mathematical clarification of spacetime contours, including the point that no physical system travels backward in time, is given in \textit{Appendix B} (End Matter).

\begin{figure*}[t]
\centering
\safeincludegraphics[width=1.5\columnwidth]{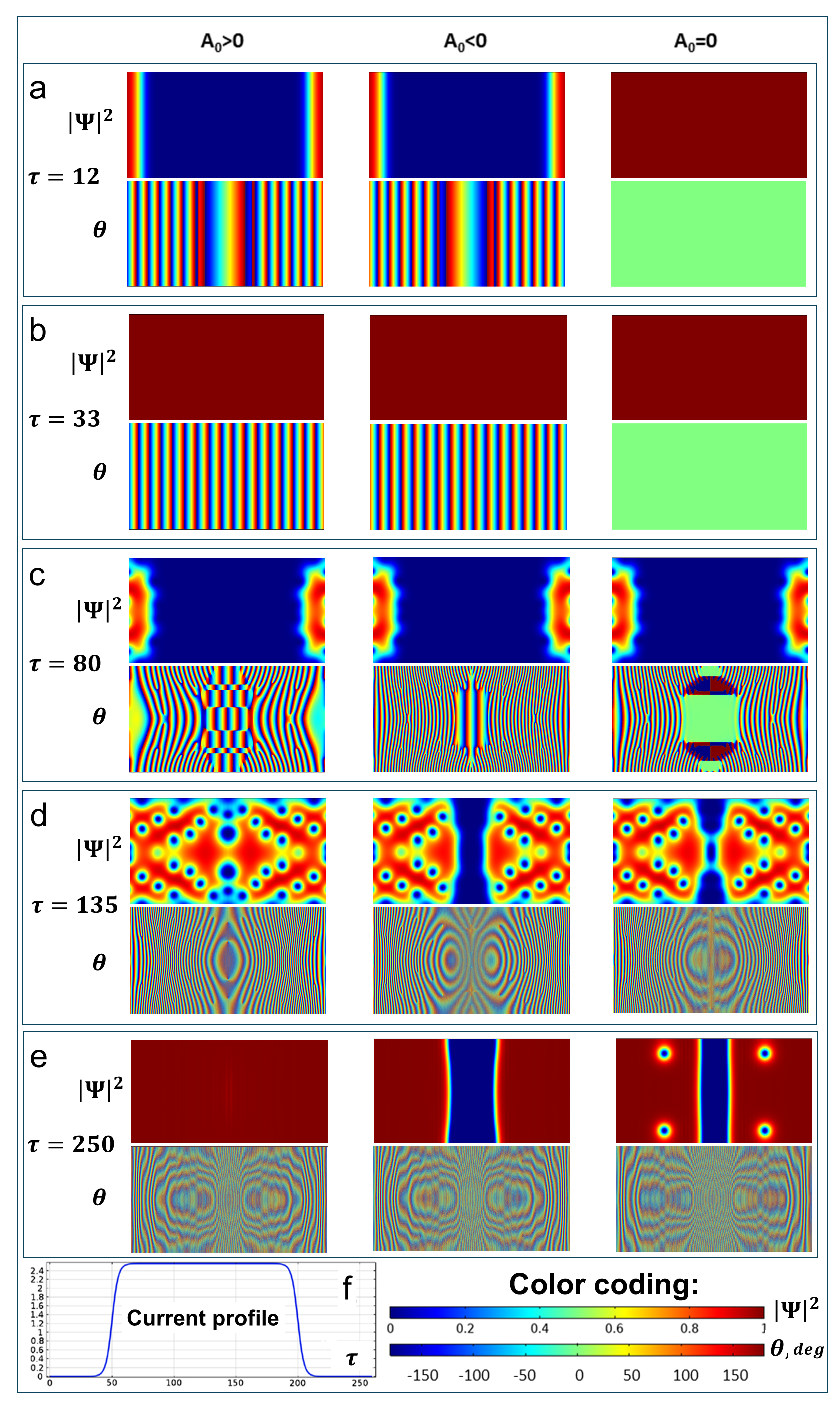}
\caption{Time evolution of condensate density $|\Psi|^2$ and its phase $\theta$ at presence of curl-free field $ \textbf{A}_0=\pm2$ and $\textbf{A}_0=0$. Small circles on the $|\Psi|^2$ plots correspond to vortices and anti-vortices. More detailed evolution is provided by an animation in the online visualization file \cite{OnlineVideo}, where annihilation of vortex-anti-vortex pairs is visible and exact timing can be seen. }
\label{fig:all panels}
\end{figure*}

In view of this understanding, we will discuss now the results which can be obtained for current-driven superconducting films in the presence of a curl-free external vector field. The outcome of the simulation based on the time-dependent Ginzburg-Landau equations \cite{Schmid1966,Gorkov1968,Golub1976,Kramer1978,Schon1979,Hu1980,WattsTobin1981,Gulian2020} is shown in Fig. \ref {fig:all panels} (a)-(e). The calculations were performed using the same model as in Ref. \cite{Gulian2018} with a refined numerical mesh and with the control case $\textbf{A}_0=0$ added to the previously studied cases $ \textbf{A}_0=\pm2$ (more details are in \textit{Appendix C}, End Matter).

At $\tau = 0$, the system begins evolving from the normal state to the superconducting state, corresponding to cooling through $T_c$. During the initial stage, no external transport current is applied, Fig.~\ref{fig:all panels} (f). Soon after nucleation begins, the condensate phase adjusts to the presence of the curl-free field $\textbf{A}_0$. The phase evolution during this interval corresponds directly to the transient voltage generation described by Eq. (9).
At $\tau = 12$, the condensate is still evolving for $A_0 \neq 0$, whereas the $A_0 = 0$ case is already essentially stationary. By $\tau \approx 25$, superconductivity is fully established and both $|\Psi|^2$ and $\theta$ become time independent ($\tau = 33$ is shown in Fig.~\ref{fig:all panels} (b)). The phase profiles are antisymmetric for $+A_0$ and $-A_0$ and uniform for $A_0 = 0$. Remarkably (see online video file [21]), the nucleation process in the presence of $A_0 = \pm 2$ is approximately five times slower than in the absence of the curl-free field.

At $\tau = 50$, a transport-current pulse is applied and maintained until $\tau = 200$, Fig.~\ref{fig:all panels} (f). The resulting vortex dynamics depend strongly on the sign and magnitude of $A_0$, leading to qualitatively different final states Fig.~\ref{fig:all panels} (d). In particular, the case $A_0 > 0$ relaxes to a fully superconducting state, whereas the other cases retain finite-resistance regions Fig.~\ref{fig:all panels} (e). 

Thus, we conclude that the nonequilibrium nucleation process already provides independent and complementary diagnostic information about the curl-free field, while the subsequent current-driven vortex dynamics further amplify and extend this nonequilibrium outcome. Together, these two regimes demonstrate that the influence of the curl-free gauge field manifests itself broadly throughout the time-dependent evolution of the superconducting system.

There are many imaginable cases where these kinds of sensors can be useful wherever closed trajectories or doubly connected arrangements are impossible. In this regard we should mention that while the ideal source of the curl-free vector potential is the infinite solenoid, similar $r$-dependence, Eq. (2), is exhibited by magnetic dipoles with high aspect ratio, say, iron needles magnetized along their axis, in which case the magnetic field amplitude decays much faster: ${H}$ $\propto 1/r^3$. This elucidates the practical importance of singly connected quantum sensors of the $\mathbf{A}$-fields.

To summarize, we have shown that the apparent topological blindness of simply connected quantum systems to curl-free gauge fields is not fundamental. During nonequilibrium evolution, quantum matter can generate measurable transient responses whose time integral is directly proportional to the open-path line integral of the vector potential. Unlike the conventional Aharonov–Bohm effect, the resulting response is not restricted modulo the flux quantum and may greatly exceed the scale associated with static doubly connected geometries. These results establish a general principle for detecting curl-free gauge fields and open new opportunities for probing hidden gauge structures in quantum matter and for developing new types of quantum sensors.

\textit{Acknowledgments}---The authors are grateful to Y. Aharonov, D. Van Vechten, D. Struppa, J. Tollaksen, G. Melkonyan, M. Gulian, and P. Abramian-Barco for numerous discussions on topics of this article.

\bigskip
The authors declare no competing interests.

\bigskip

\textit{Data Availability}---All data are available from the Contact author on reasonable request.

\begin{widetext}
\begin{center}
\textbf{End Matter}
\end{center}
\end{widetext}

The End Matter provides (A) further details on experimental observability,
(B) a clarification of spacetime Wilson-loop contours, and (C) technical
details on the application of the time-dependent Ginzburg--Landau framework.

\setcounter{section}{0}
\setcounter{equation}{0}
\renewcommand{\theequation}{\Alph{section}\arabic{equation}}
\makeatletter
\renewcommand{\theHequation}{\Alph{section}.\arabic{equation}}
\makeatother
\newcommand{\appsection}{\stepcounter{section}\setcounter{equation}{0}}

\appsection
\textit{Appendix A:~{Experimental observability}} --- The main text has emphasized the principal physical issues. For completeness we note that experimental observation of the nucleation transient would require: (i) establishment of the magnetic flux above $T_c$ so that the external field $\textbf{A}_{ext}$ is static during nucleation, (ii) sufficiently rapid cooling through $T_c$, and (iii) ultrafast voltage detection on the timescale of $\tau_{GL}$. The expected signal strength is enhanced by the fact that, unlike conventional closed loop-based measurements, the transient response scales with the absolute open-line integral of the external field rather than with the flux modulo the flux quantum. This makes wide strips and large confined flux values attractive, provided stray classical fields are adequately suppressed. For example, if the flux in the solenoid is $10^6\phi_0 $ the resulting response may exceed the conventional flux-quantized scale by many orders of magnitude.

\appsection
\textit{Appendix B:~ {Mathematical nature of spacetime gauge contours}}---The interpretation of gauge-invariant phase factors in terms of spacetime contours may give rise to a potential misunderstanding. In particular, the representation of a closed contour in 3+1 dimensional spacetime might suggest that a physical system traverses a path that includes segments propagating backward in time. It is therefore useful to clarify the mathematical meaning of such contours. The spacetime contour entering the Wilson-loop expression is a mathematical construction for phase comparison and does not imply any physical propagation of the system backward in time.

The relevant quantity in gauge theory is the phase factor associated with a path $\mathcal{C}$ in spacetime,
\begin{equation}
W(\mathcal{C})=\exp\!\left( i\int_{\mathcal{C}}A_\mu\,dx^\mu \right),
\label{eq:B1}
\end{equation}
where the path is an ordered sequence of spacetime points,
\begin{equation}
x^\mu=x^\mu(s),\qquad s_0\le s\le s_1.
\end{equation}
The parameter $s$ is not physical time; it is an auxiliary ordering variable. A closed contour satisfies $x^\mu(s_0)=x^\mu(s_1)$, but this does not mean that a physical particle or excitation traverses the contour in real time.  One may separate spatial and temporal contributions according to
\begin{equation}
\int_{\mathcal{C}}A_\mu dx^\mu=\int_{\mathcal{C}}(\mathbf{A}\cdot d\mathbf{l}-\varphi\,dt).
\end{equation}
Segments with reversed temporal orientation simply contribute with the appropriate sign. In the present context, the superconducting system does not physically travel around a spacetime loop; rather, the protocol-defined history of preparation, evolution, and readout supplies the contour needed for a gauge-invariant phase comparison.

\appsection
\textit{Appendix C: ~{Technical details of the \TDGL{} framework}}---The numerical analysis referred to in the main text is based on time-dependent Ginzburg--Landau equations in the standard gapless limit~\cite{Schmid1966,Gorkov1968,Golub1976,Kramer1978,Schon1979,Hu1980,WattsTobin1981,Gulian2020}. A convenient dimensionless form is
\begin{equation}
\left(\frac{\partial}{\partial \tau}+i\varphi\right)\Psi=-\left(\frac{i}{\kappa}\nabla+\mathbf{A}\right)^2\Psi+(1-|\Psi|^2)\Psi,
\end{equation}
accompanied by the corresponding current equation. Here $\kappa$ is the Ginzburg-Landau parameter, and, most importantly,  
\begin{equation}
\textbf{A}=\textbf{A}_{ext}+\textbf{A}_{int},
\end{equation}
where $\textbf{A}_{int}$ is the vector-potential corresponding to the magnetic field $\mathbf{H}=\nabla\times\mathbf{A_{int}}$ associated with the flow of the applied current with a profile shown in Fig. \ref{fig:all panels} (f).

Detailed COMSOL code with TDGL equations implementation, normalization, and boundary conditions were given in the previous simulation work \cite{Gulian2018}. Our additional consideration for illustrative purposes just requires applying an external field $\textbf{A}_{ext}=0$ instead of $\textbf{A}_{ext}=\pm2$ which is straightforward. Characteristic features are shown in Fig. \ref{fig:all panels}. More details can be viewed in the online video file \cite{OnlineVideo}.

\end{document}